%% file: mmse_multibeam_satellites.tex
\documentclass[journal,a4paper]{IEEEtran}
\IEEEoverridecommandlockouts

\usepackage[T1]{fontenc}
\usepackage[english]{babel}
\usepackage[utf8]{inputenc}
\usepackage{amsfonts,amsmath,amsthm,amssymb}
\usepackage{cancel}

\usepackage{graphicx}
\usepackage{float}

\usepackage{subfigure}
\usepackage{cite}
\newfloat{longequation}{b}{ext}
\newfloat{longequation}{t}{ext}
\graphicspath{{./figs/}}
\input{user_cmds}

\title{MMSE Performance Analysis of Generalized Multibeam Satellite Channels}

\author{Dimitrios Christopoulos, \IEEEmembership{Student Member, IEEE,} Jesús Arnau, \IEEEmembership{Student Member, IEEE,} \\Symeon Chatzinotas, \IEEEmembership{Member, IEEE,} Carlos Mosquera, \IEEEmembership{Member, IEEE} \\and Bj\"orn Ottersten, \IEEEmembership{Fellow, IEEE}
\thanks{The work of D. Christopoulos, S. Chatzinotas and B. Ottersten is partially supported by the National Research Fund, Luxembourg under the project  ``$CO^{2}SAT:$ Cooperative \& Cognitive Architectures for Satellite Networks'. The work of C. Mosquera is partially funded by the Seventh Framework Program project BATS (grant 3117533).}
\thanks{D. Christopoulos, S. Chatzinotas and B. Ottersten are with the SnT-University of Luxembourg. B. Ottertsen is also with the Royal Institute of Technology (KTH), Sweden.}
\thanks{J. Arnau and C. Mosquera are with the Signal Theory and Communications Department, University of Vigo, Spain. C. Mosquera is also with the Galician Research and Development center in Advanced Telecommunications (GRADIANT), Vigo, Spain.}
}
\makeindex
\usepackage{varioref}
\usepackage[normalem]{ulem}

\begin{document}

\maketitle

\begin{abstract}
 Aggressive frequency reuse in the return link (RL) of multibeam satellite communications (SatComs) is crucial towards the implementation of next generation, interactive satellite services. In this direction, multiuser detection has shown great potential in mitigating the increased intrasystem interferences, induced by a tight spectrum reuse. Herein we present an analytic framework to describe the linear Minimum Mean Square Error (MMSE) performance of multiuser channels that exhibit full receive correlation: an inherent attribute of the RL of multibeam SatComs. Analytic, tight approximations on the MMSE performance are proposed for cases where closed form solutions are not available in the existing literature. The proposed framework is generic, thus providing a generalized solution straightforwardly extendable to various  fading models over channels that exhibit full receive correlation. Simulation results are provided to show the tightness of the proposed approximation with respect to the available transmit power.
\end{abstract}

\begin{IEEEkeywords}
Multiuser detection, multibeam satellites, return link, linear minimum mean square error receivers.
\end{IEEEkeywords}

\section{Introduction \& Related Work}\label{sec:intro}
 Towards next generation broadband,  interactive, multibeam satellite systems, higher throughput requirements especially for the return link (RL) are necessary. Multi user detection (MUD) has already proven a key technique for the enhancement of the spectral efficiency of the the  RL \cite{Letzepis2008} and the Forward Link (FL) \cite{Grotz2010}   of multibeam satellite communications (SatComs). However,  the inherent differences between the satellite and the terrestrial channel impose difficulties in the theoretical performance analysis of  MUD for satellite systems.
Added to that, linear MMSE receivers are more difficult to analyze compared to the optimal capacity achieving successive interference cancellation (SIC) techniques.

 The basic characteristic of the RL multibeam satellite channel is the high receive correlation among  the channels at the satellite side, resulting from the total lack of scatterers, as well as the practical collocation of the multibeam antenna feeds,  at the satellite side. As a result, all the receive antennas experience identical fading instances from a specific user. Specifically, each user's vector channel towards the satellite antennas is defined by the antenna radiation pattern. According to the multiplicative model and assuming block fading, the product between a single fading coefficient and the channel vector will characterize the fading instance. Subsequently, the channel matrix is given by  the product of a full rank  multibeam gain matrix with a diagonal matrix, composed of random fading coefficients, as originally proposed in \cite{Letzepis2008} and extended in \cite{Christopoulos2011} and \cite{arnau2012asms}, for more realistic system models.

Linear multiuser receivers where extensively examined in\cite{Tse1999}.  More recently, numerous contributions examined the asymptotic performance of these architectures by employing random matrix theory methods  described in \cite{Tulino04} and the references therein. In the same field, the work of \cite{Kumar2009} is also noted.

Moreover, the mutual information of channels for which exact analytical formulas do not exist in literature, has been described via analytical, tight bounds. To this end, the Minkowski's inequality was initially proposed in \cite{Oyman2003} to provide a tight lower bound on the channel mutual information. This technique was extended in \cite{Matthaiou2011} to provide a new lower bound on the ergodic capacity of distributed MIMO systems. Furthermore, generic bounds on the sum rate of zero forcing (ZF) receivers over Rayleigh, Rice and correlated channels were given in \cite{Matthaiou2011b}.

Focusing on SatComs, the performance of linear MMSE receivers is not trivial to characterize analytically. In the existing literature, two different modeling approaches have been considered.  The authors of \cite{Alfano2010}  have modeled fading as the sum of two random matrices, each following a specific distribution, without however reporting results for linear receivers. The other approach, namely the multiplicative model, was introduced in \cite{Letzepis2008}, where the performance of optimal non-linear and linear receivers was given as a function of  matrix arguments, with high computational complexity,  for theoretical channel models.  This model has been extended in \cite{Christopoulos2011}, for composite, realistic channels and a novel, simple lower bound   for the channel capacity has been deduced. Finally, the incorporation of rain fading has been tackled in \cite{arnau2012asms}, using an identical channel model, in which asymptotic closed form expressions for the system's capacity and the MMSE were derived.

The purpose of the present contribution is to extend the results of \cite{Christopoulos2011} and \cite{arnau2012asms} for the case of linear MMSE receivers, thus providing  an analytic framework that covers the generalized type of channels with full receive correlation.
 To this end, the information theoretic link between the channel mutual information and the MMSE  given in  \cite{Guo2005}, is exploited\footnote{For completeness we also denote the work of \cite{McKay2010} that provides an analytic framework for linear MMSE receivers, which however does not cover our case. }. Results on the MMSE performance of the generalized satellite channel, for which the mutual information has been analytically described only through tight bounds, are proposed.




The rest of this paper is structured as follows. In Sec. \ref{sec:signal_model} the generalized multibeam satellite channel is introduced. The analytic framework proposed is described in detail in Sec. \ref{sec: analysis}. Simulation results are presented in Sec. \ref{sec:sim_results}, while Sec. concludes the findings of this paper.

\textit{Notation:}
Throughout the paper, \(\mathcal{E}\{ \cdot\}\),
\(\left(\cdot\right)^\dag\),
 denote the expectation,
 and the conjugate transpose of a matrix. Bold face lower case characters denote column vectors and upper case denote matrices,  while
$\mathbf{I}_n$ denotes an identity matrix of size $n$.

\section{Signal \& Channel Model}
\label{sec:signal_model}
Let us consider a multi-user (MU) single input multiple output (SIMO) multiple access channel (MAC) with $K$ single-antenna terminals transmitting towards a single receiver equipped with $N =K$ antennas\footnote{The commonly adopted assumption of symmetric systems (e.g. \cite{Matthaiou2011}) is a requisite for the analysis. In multibeam SatComs, it models systems that  schedule a single user per beam, each time slot. This assumption is usually  adopted in the relevant literature \cite{Letzepis2008,Christopoulos2011,Christopoulos2012}. }. The input-output relationship reads as
\begin{equation}
 \y = \sqrt{\gamma}\H\s+\n,
\end{equation}
where $\s \in \mathcal{C}^{K\times 1}$ is the transmitted baseband signal vector, such that $\mathcal E\{\s\s^H\}=\I_K$, $\y \in \mathcal{C}^{N\times 1}$ is the received signal vector, $\n\sim\mathcal{C}\mathcal{N}\left(\0,\I\right)$ is the complex noise vector and $\gamma$ is the normalized transmit signal-to-noise ratio (SNR). Also, the matrix $\H \in \mathcal{C}^{N\times K}$ represents the complex-valued block fading channel and admits a number of expressions depending on the model.

\textit{A. Generalized Multibeam Channel:}
In the present contribution, the channel matrix is assumed to be modeled as the product of a full rank matrix with a random fading matrix. The assumption of full receive correlation, that will be hereafter considered, shall impose a diagonal fading matrix\cite{Christopoulos2011,Letzepis2008} on the multiplicative model, thus yielding 
\begin{equation}\label{eq: deterministic(x)diagonal}
 \H = \B\D^{1/2},
\end{equation}
where  $\B$ is a fixed, deterministic, full rank matrix with normalized power (i.e. $\trace(\B^2)=K$)  and $\D = \diag(\d)$ is a random diagonal matrix composed of the fading coefficients.
The deterministic matrix $\B$ models the multibeam antenna radiation pattern,  while the random elements of the diagonal matrix $\D$, are drawn from distributions that model various small or large scale effects. Subsequently, the adopted model, provides a generalized approach towards systems with full receive side correlation. By changing the fading distribution different systems can be modeled. Two specific examples, relevant for multibeam SatComs will be  studied hereafter.
\subsubsection{Composite fading} This channel has been proposed in \cite{Christopoulos2011} to model the multibeam mobile satellite channel, by incorporating small scale Rician fading and large scale lognormal shadowing. The resulting model reads as
\begin{equation}\label{eq: composite channel}
 \H_{c} = \B\H_d\sqrt{\X_d},
\end{equation}\
where $\H_d$ and  $\X_d$ are diagonal fading matrices with elements drawn from Rice and log-normal distributions respectively.
\subsubsection{Rain fading} When fixed satellite services are assumed, then the high antenna directivity imposes an AWGN channel and allows for the utilization of higher frequency bands, where rain fading can dramatically deteriorate system performance. For this case the model proposed in \cite{arnau2012asms} will yield
\begin{equation}\label{eq: rain channel}
 \H_{r} = \B \sqrt{\L_{d}},
\end{equation}
where the random elements of $\L_{d}$ are drawn from a log-normal distribution in dB scale\footnote{In natural units, these elements are produced by exponentiated log-normal elements \cite{Panagopoulos2005b} and the relevant distribution is also referred to as log-log-normal.}.

\section{Analysis}
\label{sec: analysis}
The performance of linear multiuser detection (MUD) is evaluated via the achievable SINR$_k$ after MMSE detection at the $k$th user, given by \cite{Verdu1998}
\begin{equation}\label{eq: SINR mmse}
 \gamma_{\text{mmse}, k} = \frac{1}{\left[\left(\I_{K}+\gamma\H^\dag\H\right)^{-1}\right]_{kk}}-1.
\end{equation}
 Averaged over the users and all channel instances, the  system spectral efficiency,  will be given by \cite{Verdu1998}
 \begin{equation}\label{eq: mmse spef}
 \mathrm C_{\text{mmse}} =\mathcal E_\H\bigg\{\frac{1}{K}\sum_{i=1}^{K}\log_2\left(1+ \gamma_{\text{mmse}}, k\right)\bigg\},
\end{equation}
 in bits/sec/Hz. By combining \eqref{eq: SINR mmse} and \eqref{eq: mmse spef} and  directly applying the Jensen's inequality the following stands

\begin{equation}\label{eq: mmse spef bound}
 \mathrm C_{\text{mmse}}  \geq \mathcal E_\H\left\{-\log_2 \left( \frac{1}{K}\trace\left(\left(\I_{K}+\gamma\H^\dag\H\right)^{-1}\right)\right)^{}\right\}.
\end{equation}
Thus, in common practice, another measure, namely the instantaneous,  per user, minimum mean square error,   is also adopted \cite{Guo2005,Tulino04}:
\begin{equation}\label{eq:avg mmse}
\epsilon^2=\frac{1}{K}\trace\left(\left(\I_{K}+\gamma\H^\dag\H\right)^{-1}\right ).
\end{equation}
%

The presence of the inverse of a matrix sum in (\ref{eq: mmse spef bound}) hinders the computation of $\epsilon^2$ when the eigenvalue distribution of $\H^\dag\H$ cannot be computed analytically. To solve this problem, we propose to use the approximation that follows.

\begin{theorem}
The per user  MMSE of an uplink MU SIMO system can be approximated by
\begin{equation}\label{eq: main_result}
  \hat\epsilon^2 = \frac{1}{1+\gamma  \exp\left(\frac{1}{K}\ln\det\left(\H^\dag\H\right)\right)},
\end{equation}
relative to a specific channel instance.
\end{theorem}
\begin{IEEEproof}
In \cite{Guo2005}, an explicit relationship between the channel mutual information  $\mathrm{I}_e= \log_{2}\det\left(\I+\gamma\H^\dag\H\right)$ in bits/sec/Hz and $\epsilon^2$ was derived by  differentiating with respect to the SNR. This result is easily extended for vector channels  yielding\cite{Tulino04}
\begin{align}\label{eq:dif_rel}
 \gamma\frac{\partial}{\partial\gamma}\mathrm{I}_{e}(\gamma) &= K-\trace\left(\left(\I_K+\gamma\H^\dag\H\right)^{-1}\right)\\ &=K(1-\epsilon^2).
\end{align}
Furthermore, with  the use of  Minkowski's inequality, a bound on $\mathrm I_e$, tight over the whole SNR region and exact for the high SNRs, has been derived in \cite{Oyman2003}:
\begin{equation}\label{eq:minkowski}
 \mathrm{I}_e \geq \mathrm{I}_{lb}\ =
  K\log_2\bigg(1+\gamma\det\left(\left(\H^\dag\H\right)^{1/K}\right)\bigg).
\end{equation}
By differentiation with respect to  $\gamma$   we get
 the following:
\begin{align}
  \gamma\frac{\partial}{\partial\gamma}\mathrm{I}_{lb}&=  \frac{\gamma K\det\left(\left(\H^\dag\H\right)^{1/K}\right)}{1+\gamma \det\left(\left(\H^\dag\H\right)^{1/K}\right)}\stackrel{(10)}{\Longrightarrow} \nonumber\\
 \hat \epsilon^2 &= \frac{1}{1+\gamma \det\left(\left(\H^\dag\H\right)^{1/K}\right)}.\label{eq: mmse approx det}
 \end{align}
Finally  \eqref{eq: mmse approx det} can be rewritten as in \eqref{eq: main_result}.
\end{IEEEproof}
Since the differentiation does not preserve the direction of the bound, the characteristics of this approximation will be studied in more detail in the following lemma.
\begin{theorem}
 Let $\hat\epsilon^2$ be the derived approximation of $\epsilon^2$, then $\hat\epsilon^2$ and $\epsilon^2$ as functions of $\gamma$, will have a
 crossing point.
\end{theorem}

\begin{IEEEproof}
Denoting as $\lambda_i$ the ordered eigenvalues of $\H^\dag\H$,  let us define the function $D\left(\gamma\right) \stackrel{\cdot}{=} \mathrm{I}_e(\gamma)-\mathrm{I}_{lb}(\gamma)$, where $I_e\left(\gamma\right) = \sum_i\log\left(1+\gamma\lambda_i\right)$ and $I_{lb}\left(\gamma\right) = \log\left(1+\gamma \left(\prod_i \lambda_i\right)^{1/K} \right)$, with the following properties:
1) $D(0) = 0$.
2) For $\gamma_o$ sufficiently large, but finite, we have that $D(\gamma_o)= 0$.
From  properties 1 and 2, a straightforward application of Cauchy's mean-value theorem yields that there will be at least one  point  $\gamma^*\in(0,\gamma_o)$, with zero derivative and subsequently a crossing point between the approximation and the actual function.
 \end{IEEEproof}
 \begin{theorem}
 The average over the channel realizations, per user MMSE, $\mathcal E_\H\left\{\hat\epsilon^2\right\}$, can be bounded by
\begin{equation}\label{eq: main_result_exp}
   \frac{1}{1+\gamma  \exp\left(\frac{1}{K}\mathcal E_\H\bigg\{\ln\det\left(\H^\dag\H\right)\bigg\}\right)}
\end{equation}

 \end{theorem}
 \begin{IEEEproof}
 Let us consider the function $\phi(x)=\left(\left(1+\exp\left(x\right)\right)\right)^{-1}$ which is convex for $x<0$ and concave for $x>0$. By applying Jensen's inequality over the two regions we get that
\begin{equation}
 \mathcal E_\H\left\{\hat\epsilon^2\right\} =\begin{cases}\geq a,  & x\leq0 \\
\leq a, & x>0 \\
\end{cases}\
\end{equation}
for $a = \left({1+\gamma  \exp\left(\frac{1}{K}\mathcal E_\H\bigg\{\ln\det\left(\H^\dag\H\right)\bigg\}\right)}\right)^{-1}$
and

$x=1/K\cdot\ln\det(\left(\H^\dag\H\right))+\ln\gamma$.
 \end{IEEEproof}


\begin{theorem}
The average per user  MMSE  approximation, for   the composite multibeam satellite channel (as given by \eqref{eq: composite channel}), is analytically described by  \eqref{eq: approximation_comp},
 where
 $\mu_{m}$ (dB) is the mean of the normal distribution, $K_r$ the Rician factor, the $g_2$ function is given as  $g_2(s^{2})=\log s^{2} +E_i(s^2)$ (as defined in \cite{Christopoulos2011}), $E_i$ is the exponential integral and $s^2=K_{r}$ is the non-centrality parameter of the associated $\chi^2$-distribution.
\end{theorem}
\begin{IEEEproof}
 The expectation of the logarithm of the determinant of the specific channel matrix has been derived in \cite{Christopoulos2011}.
Direct application of these calculations on \eqref{eq: main_result_exp}
concludes the proof.
\end{IEEEproof}

\begin{theorem}
The per user average MMSE spectral efficiency of a multibeam satellite system under rain fading (as modeled by \eqref{eq: rain channel}) can be approximated  by the closed form formula
\begin{equation}
\mathcal E_\H\left\{ \hat\epsilon_{\text{rain}}^2\right\} = \frac{1}{1+\gamma\exp\bigg(\frac{1}{K}\ln\left(\det\B^2\right) + \mu_l\bigg)},
\label{eq: approximation_rain}
\end{equation}
 where $\mu_l $ the mean of the equivalent log-normal distribution.
\end{theorem}
\begin{IEEEproof}
Following the same line of reasoning as in the proof of Lemma 3, an analytical  bound on  $\mathcal E_\H\left\{\hat\epsilon_{\text{rain}}^2\right\}$  will read as
\begin{equation}
\frac{1}{1+\gamma  \exp\left(\frac{1}{K}\mathcal E_\H\left\{\ln\det\left(\B^2\L_d\right)\right\}\right)}
\end{equation}
where $\ln\det(\B^2\L_d)=\ln\left(\det\B^2\right)+\ln\det\L_d$ since the matrices are square and $ \mathcal E\left\{\ln\left(\det\L_d\right)\right\}=\sum_{i=1}^K\ln l_i=K\cdot\mu_l$. The parameter $\mu_l$ is the mean of the related log-normal distribution  given by $\mu_l=\exp(\mu_m+\sigma^2/2)$. As before, $\mu_m$ and $\sigma$ are the
mean and variance of the associated normal distribution.
Since $\B$ is a deterministic matrix, from the above analysis, \eqref{eq: approximation_rain} is deduced.  \end{IEEEproof}

\begin{longequation*}[tp]
\begin{align}
\mathcal E_\H\left\{\hat\epsilon_{\text{comp}}^2\right\}= \frac{1}{\left(1+\gamma\exp\bigg(\frac{1}{K}\log\left(\det\B^2\right) +\mu_m+g_2 \left( s^{2} \right)-\log\left(K_r+1\right)\right)\bigg)}.
\label{eq: approximation_comp}
\end{align}
\hrule
\end{longequation*}

\section{Simulation Results}
\label{sec:sim_results}
\begin{figure}
  \centering
  \includegraphics[width=1\columnwidth]{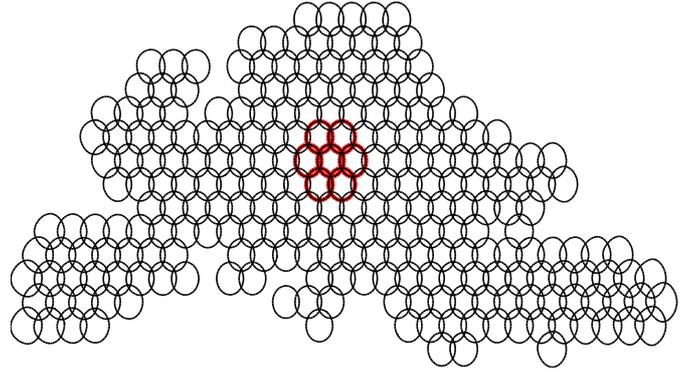}\\
  \caption{Multibeam antenna beam pattern for European coverage designed by the ESA over which a specific cluster of 7 beams (in red), is chosen. }
\label{fig: fig1}
\end{figure}
Realistic beam patterns   from multibeam coverage over Europe, that is a  245-beam pattern designed by the European Space Agency (ESA), are chosen  to provide  accurate results for the system performance (Fig. \ref{fig: fig1}).  In full-frequency multibeam systems, multiple GWs are deployed due to limited feeder link bandwidth. Each GW serves only a small number of beams.  Therefore, only a part of the total beam pattern (i.e. 7 beams) has been considered in the analysis as presented in Fig. \ref{fig: fig1}. These beams correspond to the beams served by a specific gateway (GW) over which MUD will be performed in a distributed GW scenario. Monte Carlo (MC) simulations have been performed to calculate the exact performance over the random channel. The analytic results of Sec. \ref{sec: analysis} provide deterministic approximations on the system performance with respect to $\gamma$ and are compared to the MC results to illustrate the tightness of the deduced formulas.

\begin{figure}
  \centering
  \includegraphics[width=1\columnwidth]{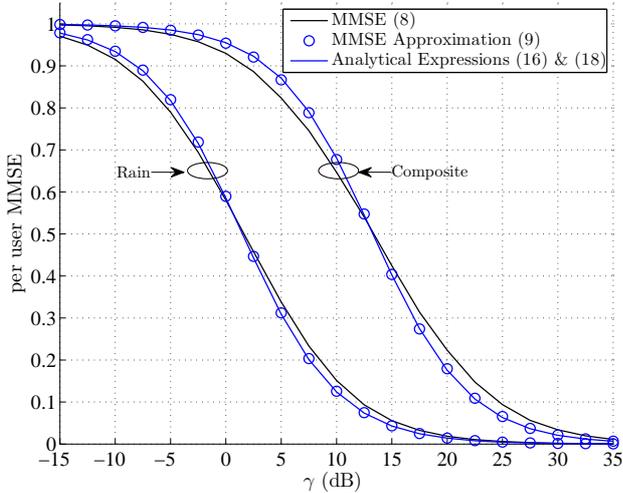}\\
  \caption{MMSE for 7 normalized ESA beams with $K_r=10$dB, $\mu_m$ = -2.63, $\sigma$=0.5.}
\label{fig: fig2}
\end{figure}

In Fig. \ref{fig: fig2}, simulation results on the expectation of the MMSE, over 1.000 channel instances of the composite and the rain faded multibeam channels, are presented. The circle markers  represent the simulated expectation over the channel instances, of the proposed approximation, given by \eqref{eq: main_result}. On top of them, the analytic formulas presented in \eqref{eq: approximation_comp} and \eqref{eq: approximation_rain}, are plotted. It is clear that these expressions precisely describe the expected values of the proposed approximation, thus providing a strong analytic tool for the description of the performance of the investigated systems. In the same figure, the tightness of the proposed approximation with respect to the actual performance, as calculated by MC simulations on \eqref{eq:avg mmse}, is also evident. Over the whole SNR region, the maximum deviation from the actual MMSE value is no more than 1.5dB for the composite fading case, while for the rain fading, the deviation is less than 1dB.
 Especially for the SNR regions around 12.5 dB and 2dB for the two cases respectively approximation becomes almost exact. In practical systems, these regions are of main interest. Consequently, the proposed expressions are very tight for systems with finite users and conventional link budgets, in contrast to asymptotic results based on large system dimensions or high SNR approaches.

\section{Conclusions}
\label{sec:conclusions}
A new, accurate approximation for the MMSE of a SIMO MU system that exhibits full receive correlation has been derived in the present paper, under an analytic framework that generalizes to various fading distributions. In particular, we have focused on the mobile composite satellite return link, as well as fixed satellite systems impaired by rain fading.
 We derive an analytic expression which approximates the expected value of the system performance in terms of MMSE, for finite system dimensions and over a large SNR region.


\nocite{Christopoulos2012}
\nocite{Christopoulos2011}

\bibliographystyle{IEEEtran}
\bibliography{refs/IEEEabrv,refs/conferences,refs/journals,refs/books,refs/references,refs/csi,refs/thesis}

\end{document}

%% file: user_cmds.tex
\def\B{{\mathbf B}}

\def\D{{\mathbf D}}

\def\H{{\mathbf H}}
\def\I{{\mathbf I}}

\def\L{{\mathbf L}}

\def\X{{\mathbf X}}

\def\0{{\mathbf 0}}
\def\1{{\mathbf 1}}

\def\d{{\mathbf d}}

\def\n{{\mathbf n}}

\def\s{{\mathbf s}}

\def\y{{\mathbf y}}

\DeclareMathOperator{\trace}{trace}

\DeclareMathOperator{\diag}{diag}

\newcommand{\openone}{\leavevmode\hbox{\small1\normalsize\kern-.33em1}}

\newtheorem{theorem}{Lemma}
